# High-energy solar particle events in cycle 24

N. Gopalswamy[1], P. Mäkelä[2,1], S. Yashiro[2,1], H. Xie[2,1], S. Akiyama[2,1], and N. Thakur[2,1]

[1]NASA Goddard Space Flight Center, Greenbelt, MD 20771, USA
[2]The Catholic University of America, Washington, DC 20064, USA

nat.gopalswamy@nasa.gov

**Abstract**. The Sun is already in the declining phase of cycle 24, but the paucity of high-energy solar energetic particle (SEP) events continues with only two ground level enhancement (GLE) events as of March 31, 2015. In an attempt to understand this, we considered all the large SEP events of cycle 24 that occurred until the end of 2014. We compared the properties of the associated CMEs with those in cycle 23. We found that the CME speeds in the sky plane were similar, but almost all those cycle-24 CMEs were halos. A significant fraction of (16%) of the frontside SEP events were associated with eruptive prominence events. CMEs associated with filament eruption events accelerate slowly and attain peak speeds beyond the typical GLE release heights. When we considered only western hemispheric events that had good connectivity to the CME nose, there were only 8 events that could be considered as GLE candidates. One turned out to be the first GLE event of cycle 24 (2012 May 17). In two events, the CMEs were very fast (>2000 km/s) but they were launched into a tenuous medium (high Alfven speed). In the remaining five events, the speeds were well below the typical GLE CME speed (~2000 km/s). Furthermore, the CMEs attained their peak speeds beyond the typical heights where GLE particles are released. We conclude that several factors contribute to the low rate of high-energy SEP events in cycle 24: (i) reduced efficiency of shock acceleration (weak heliospheric magnetic field), (ii) poor latitudinal and longitudinal connectivity), and (iii) variation in local ambient conditions (e.g., high Alfven speed).

## 1. Introduction

Solar Cycle 24 has been the weakest in the space era as measured by the sunspot number (SSN). The average SSN over the first 73 months of cycle 23 was ~ 46 in cycles 24 compared to 76 over the same epoch in cycle 23. This corresponds to a decrease of ~40% [1]. The activity has already entered into the declining phase, but the number of high-energy solar energetic particle (SEP) events has remained very low. During the first 73 months of cycle 24, there were only two ground level enhancement (GLE) events: 2012 May 17 [2] and 2014 January 6 [3]. Over the same epoch, there were 9 GLE events in cycle 23. Thus the reduction in the number of GLE events is by 78%, much higher than that in SSN. The number of >500 MeV SEP events was also higher in cycle 23 (18 vs. 6 in cycle 24).

The current paradigm for SEPs is that they are accelerated by shocks driven by coronal mass ejections (CMEs) [4]. The majority of CMEs associated with SEPs are halos and the average speed is ~1500 km/s [5] suggesting that the CMEs are very energetic. In cycle 24, the number of fast (≥900 km/s) and wide (≥60 degrees) declined only slightly (~30%), which means the reduction in the number of high-energy SEP events cannot be accounted for by the decline of the number of fast and wide CMEs alone.

It has been suggested that the weakened heliospheric magnetic field reduces the efficiency of particle acceleration leading to the lack of high-energy SEP events [6]. Another study considered CMEs associated with 59 major solar flares of X-ray class ≥M5.0 and found that only 16 of them (or 27%) were associated with large SEP events [7]. Ten other large SEP events were associated with weaker flares (soft X-ray flare size <M5.0). This study showed that the flare size is not a good indicator of SEP events. On the other hand, the CMEs associated with the SEP events were all fast, capable of driving shocks. Furthermore, a large number of fast CMEs (>2000 km/s) associated with SEP events occurred at large ecliptic distances suggesting that the nose of these CMEs were not connected to Earth, so high energy particles accelerated at the nose might not have reached Earth.

In this work, we extend the study to a larger set of SEP events, covering events up to the end of 2014. In addition, we consider a wider range of longitudes (from W00 to W90). We confine ourselves to large SEP events: >10 MeV GOES proton intensity ≥10 particles.cm$^{-2}$s$^{-1}$sr$^{-1}$. We compare CME kinematics and source positions over the corresponding epoch in cycle 23 in order to understand the multiple factors involved in the occurrence of GLE events.

## 2. Data

We make use of the list of large SEP events that are cataloged and maintained at the CDAW Data Center (http://cdaw.gsfc.nasa.gov/CME_list/sepe/) [8]. Since cycle 24 started around 1 December 2008, the study period corresponds to the first 73 months of the cycle. For comparison purposes, we use SEP events from cycle 23 over the corresponding epoch: May 1996 to May 2002. The SEP events were obtained from the NOAA web site, ftp://ftp.swpc.noaa.gov/pub/indices/SPE.txt available until the end February 2014. After this period, we examined the GOES proton data to identify additional events. We also carefully checked the events and separated some compound events. The CME information was obtained from the coronagraphs on board the Solar and Heliospheric Observatory (SOHO) and the Solar Terrestrial Relations Observatory (STEREO). In particular, we use the images from the Large Angle and Spectrometric Coronagraph (LASCO) on board SOHO [9] and the Sun Earth Connection Coronal and Heliospheric Investigation (SECCHI) coronagraphs on board STEREO [10]. For each of the SEP event, we identified the corresponding CME from the SOHO/LASCO catalog [8]. Events to the end of January 2014 were also reported in [7].

There were 37 large SEP events during the study period as listed in Table 1. Column 2 of Table 1 gives the date and time the SEP intensity in the >10 MeV channel increased above the 10 pfu level. Column 3 gives the peak SEP intensity attained. In some cases, the SEP intensity was the highest during the shock arrival at 1 AU marked by the energetic storm particle (ESP) event, but we note only the non-ESP part. For each SEP-associated CME, we identified the solar source as the heliographic coordinates of the associated flare or the centroid of the eruptive filament (column 4). For most of the SEP events, STEREO observations of the associated CMEs are also available, so we were able to fit flux ropes to the CMEs using multiple views from SOHO and STEREO [11]. This gives the flux rope location, which we correct for the solar B0 angle. The resulting final location of the flux ropes is given in column 5. From STEREO EUV images, we identified the source coordinates even for the backside events. We also note the soft X-ray flare size (column 6) and the source NOAA active region (AR) number (column 7) if available. When the source region is a quiescent filament region, we mark it as "EP" for eruptive prominence. For event #37, there were two filament eruptions, one associated with the SEP event, and the other an eastern event. The soft X-ray enhancement from the eastern event was higher, so we do not know the soft X-ray size (marked ????) for the western event we are interested in. Backside source regions are marked as "Back". Column 8 gives the first appearance time of the associated CMEs in the LASCO/C2 FOV. The sky-plane speed ($V_{sky}$) compiled from the SOHO/LASCO catalog are listed in column 9. The space speeds ($V_{sp}$) obtained from the flux rope fit are given in column 10. $V_{sp}$ is the peak space speed attained by the CMEs as derived from the flux-rope fit. For event #36, the STEREO data was not useful for fitting. For the last event, there was no STEREO data. Finally, the width ($W$) of the CMEs is given in column 11, where $H$ denotes halo CMEs. For the non-halo CMEs, the sky-plane width is given in degrees.

Table 1. List of large SEP events from cycle 24

| # | SEP Date and UT | Ip pfu | Source Loc. | Final Loc. | Flare Imp. | AR # | CME UT | $V_{sky}$ | $V_{sp}$ | W |
|---|---|---|---|---|---|---|---|---|---|---|
| 1 | 2010/08/14 12:30 | 14 | N12W56 | S20W54 | C4.4 | 11099 | 08/14 10:12 | 1205 | 1658 | H |
| 2 | 2011/03/08 01:05 | 50 | N24W59 | N39W58 | M3.7 | 11164 | 03/07 20:00 | 2125 | 2660 | H |
| 3 | 2011/03/21 19:50 | 14 | N23W129 | N33W125 | Back | 11169 | 03/21 02:24 | 1341 | 2022 | H |
| 4 | 2011/06/07 08:20 | 72 | S21W54 | S08W51 | M2.5 | 11226 | 06/07 06:49 | 1255 | 1680 | H |
| 5 | 2011/08/04 06:35 | 96 | N16W30 | N13W30 | M9.3 | 1 1261 | 08/04 04:12 | 1315 | 2450 | H |
| 6 | 2011/08/09 08:45 | 26 | N17W69 | N02W68 | X6.9 | 11263 | 08/09 08:12 | 1610 | 2496 | H |
| 7 | 2011/09/23 22:55 | 35 | N11E74 | S02E83 | X1.4 | 11302 | 09/22 10:48 | 1905 | 2474 | H |
| 8 | 2011/11/26 11:25 | 80 | N27W49 | N08W47 | C1.2 | EP | 11/26 07:12 | 933 | 1187 | H |
| 9 | 2012/01/23 05:30 | 3000 | N28W36 | N35W22 | M8.7 | 11402 | 01/23 04:12 | 2102 | 2150 | H |
| 10 | 2012/01/27 19:05 | 800 | N27W71 | N33W82 | X1.7 | 11402 | 01/27 18:27 | 2508 | 2625 | H |
| 11 | 2012/03/07 05:10 | 1500 | N17E27 | N25E31 | X5.4 | 11429 | 03/07 00:24 | 2544 | 2987 | H |
| 12 | 2012/03/13 18:10 | 500 | N19W59 | N28W52 | M7.9 | 11429 | 03/13 17:36 | 1898 | 2333 | H |
| 13 | 2012/05/17 02:10 | 255 | N11W76 | S05W76 | M5.1 | 11476 | 05/17 01:47 | 1618 | 1997 | H |
| 14 | 2012/05/27 05:35 | 14 | N10W121 | N12W115 | Back | 11482 | 05/26 20:57 | 1966 | 2623 | H |
| 15 | 2012/06/16 19:55 | 14 | S17E06 | S27E02 | M1.9 | 11504 | 06/14 14:12 | 981 | 1626 | H |
| 16 | 2012/07/07 04:00 | 25 | S18W51 | S32W62 | X1.1 | 11515 | 07/06 23:12 | 1912 | 2464 | H |
| 17 | 2012/07/09 01:20 | 18 | S17W74 | S38W88 | M6.9 | 11515 | 07/08 16:54 | 1211 | 2905 | 157 |
| 18 | 2012/07/12 18:35 | 96 | S15W01 | S23W06 | X1.4 | 11520 | 07/12 16:48 | 1360 | 1415 | H |
| 19 | 2012/07/17 17:15 | 136 | S28W65 | S32W79 | C9.9 | 11520 | 07/17 13:48 | 814 | 1881 | H |
| 20 | 2012/07/19 07:45 | 70 | S28W75 | S20W88 | M7.7 | 11520 | 07/19 05:24 | 1681 | 2048 | H |
| 21 | 2012/07/23 15:45 | 12 | S17W141 | N00W135 | Back | 11520 | 07/23 02:36 | 2142 | 2621 | H |
| 22 | 2012/09/01 13:35 | 59 | S19E42 | S13E40 | C8.4 | EP | 08/31 20:43 | 1167 | 1601 | H |
| 23 | 2012/09/28 03:00 | 28 | N06W34 | N09W29 | C3.7 | 11577 | 09/28 00:12 | 1035 | 1479 | H |
| 24 | 2013/03/16 19:40 | 16 | N11E12 | N17E08 | M1.1 | 11692 | 03/15 07:12 | 980 | 1602 | H |
| 25 | 2013/04/11 10:55 | 114 | N09E12 | N14E11 | M6.5 | 11719 | 04/11 07:24 | 986 | 1626 | H |
| 26 | 2013/05/15 13:35 | 41 | N11E51 | N15E64 | X1.2 | 11748 | 05/15 01:48 | 1511 | 2294 | H |
| 27 | 2013/05/22 14:20 | 1660 | N15W70 | N04W59 | M5.0 | 11745 | 05/22 13:32 | 1537 | 1881 | H |
| 28 | 2013/06/23 20:10 | 14 | S16E73? | ???? | M2.9 | 11777 | 06/21/03:12 | 1900 | 1987 | >207 |
| 29 | 2013/09/30 05:05 | 200 | N23W25 | N16W29 | C1.1 | EP | 09/29 22:12 | 1025 | 1864 | H |
| 30 | 2013/12/28 21:50 | 30 | S08W130 | N02W127 | Back | ???? | 12/28 17:36 | 1133 | 1918 | H |
| 31 | 2014/01/06 09:15 | 40 | S15W112 | S06W102 | Back | 11936 | 01/06 08:00 | 1402 | 2287 | H |
| 32 | 2014/01/07 19:55 | 1026 | S12W11 | S15W29 | X1.2 | 11944 | 01/07 18:24 | 1830 | 3121 | H |
| 33 | 2014/02/20 08:50 | 22 | S15W73 | S07W70 | M3.0 | 11976 | 02/20 08:00 | 948 | 1281 | H |
| 34 | 2014//02/25 13:55 | 24 | S12E82 | S10E80 | X4.9 | 11990 | 02/25 01:25 | 2147 | 2218 | H |
| 35 | 2014/04/18 15:25 | 58 | S20W34 | S18W18 | M7.3 | 12306 | 04/18 13:25 | 1203 | 1711 | H |
| 36 | 2014/09/11 02:40 | 29 | N14E02 | ???? | X1.6 | 12158 | 09/10 18:00 | 1425 | 1652 | H |
| 37 | 2014/11/02 21:10 | 11 | N27W79 | ???? | ???? | EP | 11/01 06:12 | 740 | 1000 | H |

The CME identification is unambiguous for all the events except for the 2013 June 23 event (#28 in Table 1). There was a minor particle event associated with the listed CME on June 21, but the origin of the increase on June 23 is uncertain. The increase is well after a shock observed at 1 AU, so it does not seem to be the ESP event associated with the June 21 CME. Because of this, we do not include this event in this study.

### 3. Analysis and Results

*3.1 SEP occurrence rate*

There were 37 SEP events in cycle 24 compared to 60 in cycle 23. If we normalize the number of events to the average SSN during the study period, we get 0.80/SSN in cycle 24 very similar to 0.79/SSN in cycle 23. The number of events with >100 pfu is 24 (or 40%) in cycle 23 compared to just 10 (or 27%) in cycle 24. The highest SEP intensity was 3000 pfu in cycle 24 compared to 31700 pfu in cycle 23. There were only 4 SEP events in cycle 24 with intensity >1000 pfu, compared to 10 in cycle 23. The difference is more striking when we consider >500 MeV (6 vs. 18) and GLE (2 vs. 9) events in cycle3 23 and 24. Normalized to SSN, cycle 24 had only 0.04 GLEs/SSN compared to 0.11/SSN in cycle 23.

This is in stark contrast to the >10 MeV events, which are similar when normalized to SSN. We focus on the paucity of GLE events in cycle 24.

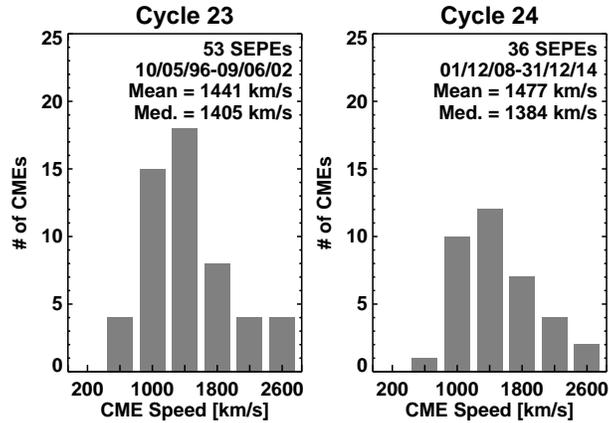

Figure 1. Sky-plane speed distributions of SEP-associated CMEs in cycles 23 (left) and 24 (right). The mean and median values of the distributions are shown on the plot.

*3.2. CME kinematics*
Figure 1 compares the speed distributions of SEP-associated CMEs in cycles 23 and 24. We are comparing sky-plane speeds because there was no 3-dimensional information in cycle 23. There were actually 60 SEP events during the first 73 months of cycle 23, but only 53 had CME data available (mainly due to SOHO/LASCO data gaps in 1998 and 1999). The distributions are very similar with almost the same mean and median speeds. The most probable value of the speed is in the 1400 km/s bin in both cycles. The range of speeds is also the same. While the speeds are similar, the width distribution is clearly different. We did not plot the width distribution because almost all CMEs are halos (35 out of 36 or 97%) in cycle 24 compared to 75% (40 out of 53) in cycle 23. Such a difference in width distribution was also found for limb CMEs in cycle 24 and was attributed to the anomalous expansion of CMEs due to the reduced total pressure in the heliosphere [6]. It was also recently reported that the occurrence rate of halo CMEs in cycle 24 did not diminish unlike SSN [1]. We note that there were five EP events in Table 1 (16% of the 31 front-side events). CMEs associated with EP events generally have a small initial acceleration and the CMEs attain high speeds at large distances from the Sun [12].

*3.3. Source locations*
Table 1 shows that the CME sources were located from close to the east limb (E82 for #34) to far behind the west limb (W141 for #21). There were 5 backside events, all of them behind the west limb. The solar sources of the remaining 31 events are plotted in Figure 2. The heliographic grid corresponds to the case of B0=0, so the latitudes are the same as the ecliptic distances (the projection of Earth is at the disk center in this representation). The source locations are the flux rope (FR) latitudes and longitudes obtained from the flux rope fit corrected for the solar B0 angle (see [7] for details). The FR locations are generally different from the flare locations due to non-radial motion of CMEs caused by magnetic pressure gradients around the eruption sites. The latitudes differ by ~10 degrees, while the longitudes differ by ~7 degrees on the average. Both the GLE CMEs have deflections that reduce the ecliptic distance. Only 10 events are in the eastern hemisphere because events from the western hemisphere are magnetically better connected to Earth. The eastern events are generally associated with more energetic CMEs and the intensities are lower. We see that the ecliptic latitudes extend to more than 30 degrees both in the northern and southern hemispheres. The latitudes of the eastern events are within 30-degree latitudes.

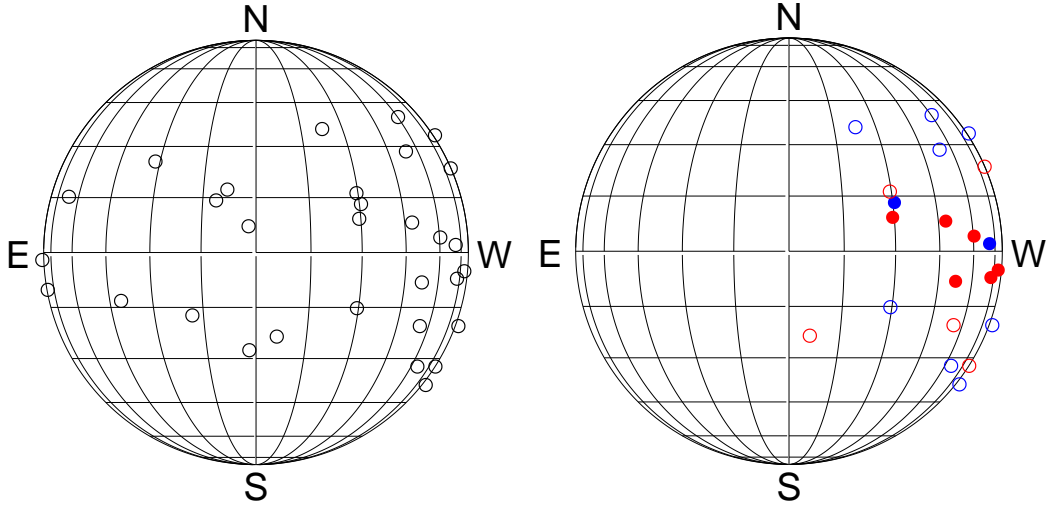

Figure 2. (left) Flux rope locations of 31 frontside SEP events from Table 1. The locations have been corrected for B0 angle (the zero latitude is the ecliptic in this figure). (right) 21 events from the western longitudes (W00 to W90). The red (blue) circles correspond to CMEs with speed < 2000 km/s (≥ 2000 km/s). The filled circles are within the ecliptic latitude range ±13 degrees. The filled circles represent the candidate GLE events discussed in the text.

Table 2. List of candidate GLE events

| # | CME Date UT | Imp. | Flr Loc. | FR Loc | B0 | Final Loc | Vsp | Max E | Ip |
|---|---|---|---|---|---|---|---|---|---|
| 4 | 2011/06/07 06:16 | M2.5 | S21W54 | S08W51 | +0.1 | S08W51 | 1680 | 330-420 | 72 |
| 5 | 2011/08/04 03:41 | M9.3 | N19W36 | N19W30 | +6.0 | N13W30 | 2450 | 165-500 | 96 |
| 6 | 2011/08/09 07:48 | X6.9 | N17W69 | N08W68 | +6.3 | N02W68 | 2496 | 330-420 | 26 |
| 8 | 2011/11/26 06:09 | C1.2 | N08W49 | N10W47 | +1.5 | N08W47 | 1187 | 40-80 | 80 |
| 13 | 2012/05/17 01:25 | M5.1 | N11W76 | S07W76 | -2.4 | S05W76 | 1997 | >700 | 255 |
| 23 | 2012/09/27 23:24 | C3.7 | N06W34 | N16W29 | +6.9 | N09W29 | 1479 | 80-165 | 28 |
| 27 | 2013/05/22 13:08 | M5.0 | N15W70 | N02W59 | -1.8 | N04W59 | 1881 | 330-420 | 1660 |
| 33 | 2014/02/20 07:26 | M3.0 | S15W73 | S14W70 | -7.0 | S07W70 | 1281 | 330-420 | 22 |

### 3.4. Analysis of GLE Candidate Events

For the purposes of this paper, we consider all the western events, as it is well known that GLEs occur preferentially from the western hemisphere [13]. The right-side plot in Figure 2 shows the 21 front-side western events, distinguishing SEP events according to their speeds: <2000 km/s (red) and ≥2000 km/s (blue). We made this distinction based on the fact that GLEs are associated with CMEs with an average speed of ~2000 km/s [14]. We have also isolated SEP events with ecliptic latitudes within ±13 degrees (filled circles). This is based on the observation that the average source latitude of cycle-23 GLE events was 13 degrees [2]. The two GLE events of cycle 24 are well within this 13-degree limit (#13 and #31 in Table 1). Applying all these criteria, we arrive at just 8 events that are in the right latitude range. We have listed the CMEs associated with these 8 events in Table 2 along with the flux rope location, B0 angle, and the final location. We now discuss these events in an attempt to explain why they did not have the GLE component except for the 17 May 2012 event.

*3.4.1. The three fastest events.* Among the three fastest CMEs in Table 2, only one was associated with a GLE (17 May 2012 event), discussed extensively in [2]. The CME speed (1997 km/s) was just at the average value of GLE-associated CMEs. The two >2000 km/s CMEs did not produce GLEs. The 9 August 2011 event was reported to be associated with narrower CME launched into a tenuous corona with probably higher than usual Alfven speed in the ambient medium (hence weaker shock – see [2]). The 4 August 2011 event was not discussed in [2] because a stricter latitude criterion (±9 degrees) was used based on cycle-23 GLE events that occurred closer to the limb. The 4 August 2011 event occurred at N13W30 just at the latitude limit for GLEs. The W30 longitude in Earth view does not allow us to examine the nature of the corona. Fortunately, the CME was observed by STEREO-Ahead as an east limb event (E70 in that view because the spacecraft was at W100) as can be seen in Figure 3. In the STEREO-Ahead COR2 image taken at 04:24 UT, close to the onset time of the 4 August 2011 SEP event at Earth, we see that the corona overlying the CME is the darkest compared to other parts of the image, suggesting that the corona was tenuous. Although we do not have magnetic field measurement, we speculate that the Alfven speed is higher than usual leading to a weaker shock as in the case of the 9 August 2011 event.

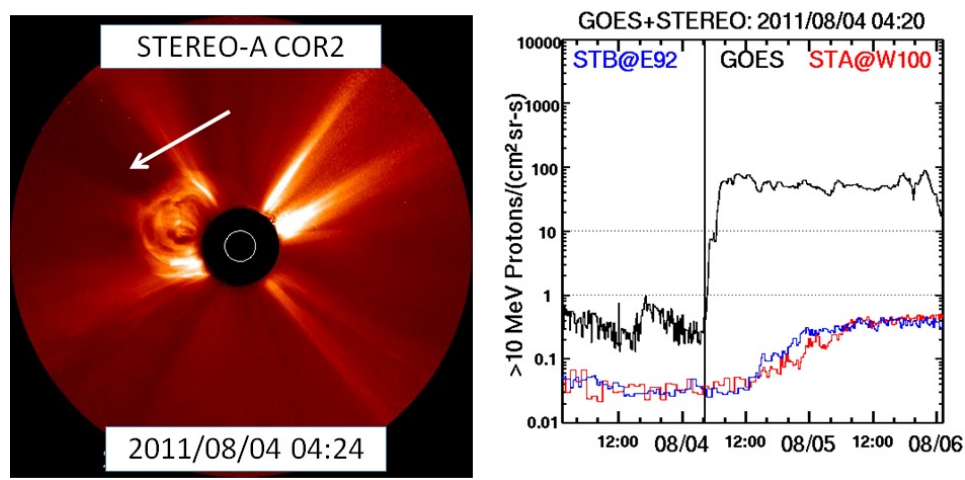

Figure 3. STEREO-Ahead COR2 image (left) showing the 4 August 2011 CME above the east limb. The arrow points to the dim corona above the CME. The >10 MeV proton intensity from GOES and STEREO-Ahead and Behind are shown for comparison. Since the two STEREO spacecraft are poorly connected to the source region, they detected only minor events (<1 pfu).

*3.4.2. The two events associated with C-class flares.* The 26 November 2011 and the 27 September 2012 CMEs were both associated with C-class flares and discussed in [12]. The 26 November 2011 CME was associated with a filament eruption. Both events were associated with interplanetary type II bursts, but no metric type II bursts. This means the shock forms only in the outer corona, well above the typical height of GLE particle release. From the flux rope fit, we estimate that the shock formed at a height of 4 solar radii (Rs) for the 27 September 2012 CME and 3.25 Rs for the 26 November 2011 CME. In addition, the peak SEP spectra were steep (index = 4.34 for the 26 November 2011 event and 3.25 for the 27 September 2012 event) [15]. The steep spectrum is also evident from the maximum energy (E) of protons detected in GOES differential channels (80 and 165 MeV for the 26 November 2011 and 27 September 2012 CMEs – see Table 2). Furthermore, these CMEs were still accelerating in the inner corona, so the peak speeds listed in Table 2 were not achieved until after the shock formation heights. Therefore, we conclude that in these two events the speed was too low and the shock formed well beyond the typical GLE shock heights (2–3 Rs).

*3.4.3 The remaining three events.* The 7 June 2011 and 20 February 2014 CMEs had peak speeds well below the typical GLE speeds (1680 and 1281 km/s, respectively). The 22 May 2013 CME was interacting with a preceding CME, but the interaction was taking place at much larger height (~20 Rs) [7]. Even though CME interaction events have high intensity (which is true in this event also) [5], the shocks need to be much closer to the Sun to be relevant for GLE events. Furthermore, the peak speed was only 1881 km/s, not quite the average value of GLE CMEs. The peak speed was also attained when the CME was at ~7 Rs. The metric type II burst started only around 40 MHz, which is a good indication that the shock forms at a larger height (~2 Rs). The typical shock formation heights for GLE events is ~1.5 Rs [2, 16]. The combination of low speed and larger shock formation height seem to have contributed to the lack of GLEs in the 22 May 2013 event. GLEs have been reported to be associated with much lower speeds when preconditioning was present [17]. The 22 May 2013 event seems to be such a case, yet there was no GLE. The overall reduced efficiency of particle acceleration might have also contributed to the paucity of GLEs in almost all the SEP events of cycle 24.

## 4. Discussion

The above analysis shows that there are three primary factors that affect the acceleration of SEPs to the highest energies: (i) CME speed, (ii) connectivity, and (iii) ambient conditions. The lowest space speed in Table 1 is 1187 km/s, which is definitely high enough to accelerate SEPs. However, such speeds may not be sufficient to make the shock strong enough to accelerate particles to the highest energies. This is compounded by the reduced ambient magnetic field strength in cycle 24 [6,18] that reduces the acceleration efficiency of shocks. For GLEs, the highest CME speed is attained close to the Sun, so the shock is in the high magnetic field region for a longer time. In some cases, the speed was relatively high, but was attained at a larger heliocentric distance. This is clear in filament eruption events and in events where the metric type II bursts start at low frequencies.

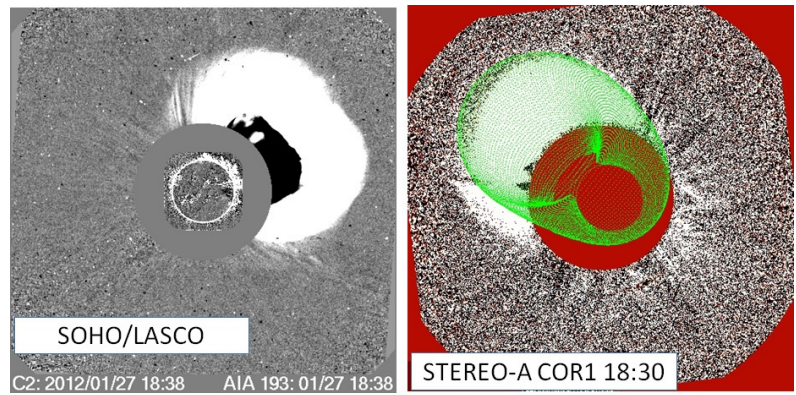

Figure 4. (left) SOHO/LASCO CME on 27 January 2012 at 18:38 UT associated with an intense SEP event. (right) STEREO-A COR1 image at 18:30 UT with the fitted flux rope (green contours) superposed. The CME is heading to the northwest in Earth view and northeast in STEREO-A view because the latter was at W108.

Under connectivity, we considered western events because it is well known that most of the GLE events originate in the western hemisphere. We also considered latitudinal connectivity [19], by which we mean that the shock nose needs to be close to the ecliptic (within 13 degrees based on cycle 23 GLE events. This may suggest that the highest energy particles may be accelerated near the nose, but they do not reach an Earth observer because the nose is too far away from the ecliptic. This is illustrated in Figure 4 for the 27 January 2012 SEP event. The flare location was N27W71 hence longitudinally well connected, but the latitude was very high. The nose of the LASCO CME was at position angle 296 degrees, consistent with the N27 latitude. The flux rope fit in Figure 4 is consistent with the higher latitude of the source region. The fitted longitude was slightly higher (W82) [7] but the latitude was the same as that of the flare location (event # 10 in Table 1). In addition, the B0 angle was -5.6 degrees, making the effective ecliptic latitude as N32.6 and hence worsening the latitudinal connectivity. This is well beyond the average ecliptic distance of 13 degrees for GLE events. From Figure 4 we can infer that only the southern flank of the CME is at the ecliptic. The SEP event was intense (800 pfu) and the GOES differential channel data indicate energetic particles present up to the 510–700 MeV channel [7]. Thus this is an almost GLE event, but not detected by the neutron monitors as a GLE.

In the list of GLE candidates in Table 2, we considered only front-side events. However, we do have the 2014 January 6 GLE event originating from behind the west limb (event #31 in Table 1). Furthermore, there was a GLE in cycle 23 on 2001 April 18 from similar longitudes (W117, see [14]). Therefore, we briefly discuss the possibility of GLEs from behind-the-west-limb events in Table 1. There are 5 such backside events, including the 2014 January 6 GLE. Of these, the 2012 May 27 event is closest in longitude to the limb (W115 or ~ 25 degrees behind the limb) and the ecliptic distance is only N12. This makes it a viable GLE candidate. Other events had longitudes in the range W125 – W135, too far for magnetic connection to Earth. Figure 5 compares the post eruption arcades (PEAs) of the 2012 May 27 and 2014 January 6 events as observed by the STEREO-A EUVI instrument. The PEAs are oriented differently in the two cases (January 2014: east-west and May 2012: north-south). These images suggest that the flux ropes from these source regions have different inclinations. The fitted flux ropes are consistent with this expectation (not shown). The shock surrounding a high inclination flux rope would have a smaller longitudinal extent than that surrounding a low inclination flux rope. We suggest this as the possible reason for not observing GLEs from the 2012 May 27 SEP event because the shock does not extend enough in the longitudinal direction. The different flux rope orientations may also result in different longitudinal extents of SEP events [20].

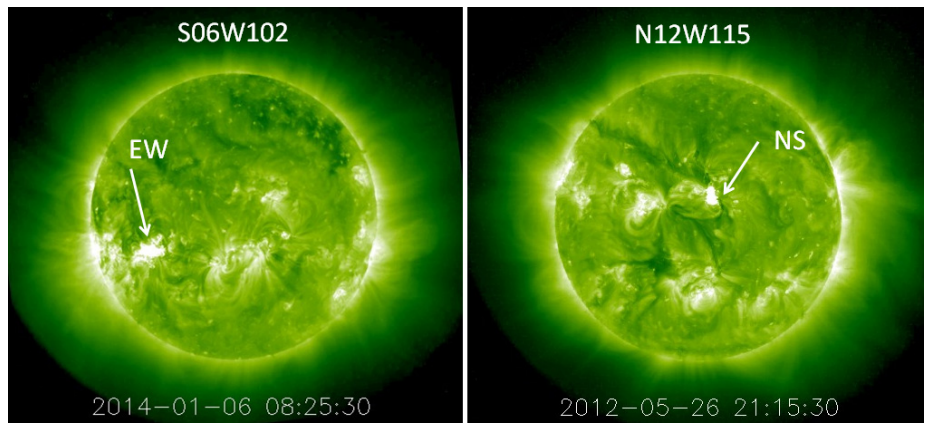

Figure 5. Post eruption arcades (PEAs) associated with the 2014 January 6 (left) and the 2012 May 27 (right) backside SEP events as observed by the Extreme Ultraviolet Imager (EUVI) on board STEREO-A. STEREO-A was located at W151 and W115 when the January 2014 and May 2012 images were obtained. Note the different orientations of the PEAs: east-west (EW) and north-south (NS).

We also considered the possibility of higher Alfven speed in the ambient medium indicated by the tenuous corona. This is qualitative, but can be verified by estimating the coronal magnetic field [21]. Interestingly, most of the >2000 km/s CMEs have higher ecliptic latitudes. It is not clear if this is due to the overall heliospheric magnetic field structure in cycle-24.

**5. Summary**
We confirm our previous finding that the number of >10 MeV SEP events relative to SSN is similar in cycles 23 and 24, but there were four times more GLE events in cycle 23 than those in cycle 24 during the first 73 months. When normalized to SSN, the rates are 0.04/SSN in cycle 24 compared to 0.11/SSN in cycle 23. The speed distributions of CMEs associated with SEPs are quite similar in the two cycles with little difference in the mean and median speeds. However, the fraction of halos is 97% in cycle 24 compared to 75% in cycle 23. The higher fraction of halos was also reported earlier for a set of limb CMEs associated with soft X-ray flares of size ≥C3.0 [6]. The increased expansion of cycle-24 CMEs seems to have contributed to the increased number of halo CMEs in cycle 23. Since GLE events are a subset of SEP events, we considered front-side SEP events in the western hemisphere as the GLE candidate events. Based on the recent result that the ecliptic distance of CMEs needs to be small, we further narrowed down the events to ecliptic distances ≤13 degrees. This resulted in 8 candidate events including the 17 May 2012 GLE. Among the 7 non-GLE events, two events had very high speed (>2000 km/s), but the ambient medium was not favorable for supporting strong shocks (probably high Alfven speed). Four of the five remaining events had speeds well below the average speed of GLE CMEs (~2000 km/s). The last event (22 May 2013) was an interacting CME event but the interaction distance was ~20 Rs, well beyond the distance where GLE particles are typically released. Furthermore, the peak speed was attained around 7 Rs. Thus we conclude that the CME speed, connectivity of CME nose to Earth, and favorable ambient medium are the key factors that determine whether an SEP event has a GLE component.

**Acknowledgments**
SOHO is a project of international collaboration between ESA and NASA. STEREO is a mission in NASA's Solar Terrestrial Probes program. The work of NG, SY, SA was supported by NASA/LWS program. PM was partially supported by NSF grant AGS-1358274 and NASA grant NNX15AB77G. HX was partially supported by NASA grant NNX15AB70G.